\documentclass{aa}  
\usepackage{graphicx}
\usepackage{natbib}
\usepackage{txfonts}
\usepackage[colorlinks=true,citecolor=blue]{hyperref}
\usepackage[normalem]{ulem}
\begin{document}

   \title{Asteroseismology of evolved stars to constrain the internal transport of angular momentum}

   \subtitle{VI. Testing a parametric formulation for the azimuthal magneto-rotational instability}

      \author{F.D. Moyano
          \inst{1}
          \and
          P. Eggenberger\inst{1}
          \and
          B. Mosser \inst{2}
          \and
          F. Spada \inst{3}
      }
          
          \institute{Observatoire de Gen\`eve, Universit\'e de Gen\`eve, 51 Ch. Pegasi, CH-1290 Versoix, Suisse \\
            \email{facundo.moyano@unige.ch}
            \and LESIA, Observatoire de Paris, Universit\'e PSL, CNRS, Sorbonne Universit\'e, Universit\'e de Paris, 92195 Meudon, France
            \and Max-Planck-Institut für Sonnensystemforschung, Justus-von-Liebig-Weg 3, 37077, Göttingen, Germany
          }

          \date{Received --; accepted 13 Feb 2023}
          \titlerunning{Angular momentum transport by the AMRI}

% \abstract{}{}{}{}{} 
% 5 {} token are mandatory
 
  \abstract
      {Asteroseismic measurements of the internal rotation rate in evolved stars pointed out to a lack of angular momentum (AM) transport in stellar evolution models.
        Several physical processes in addition to hydrodynamical ones were proposed as candidates for the missing mechanism.
        Nonetheless, no current candidate can satisfy all the constraints provided by asteroseismology.}
      {We revisit the role of a candidate process whose efficiency scales with the contrast between the rotation rate of the core and the surface which was proposed to be related to the azimuthal magneto-rotational instability (AMRI) by Spada et al.}
   {We compute stellar evolution models of low- and intermediate-mass stars with the parametric formulation of AM transport proposed by Spada et al. until the end of the core-helium burning for low- and intermediate-mass stars and compare our results to the latest asteroseismic constraints available in the post main sequence phase.}
   {Both hydrogen-shell burning stars in the red giant branch and core-helium burning stars of low- and intermediate-mass in the mass range $1 M_{\odot} \lesssim M \lesssim 2.5 M_{\odot}$ can be simultaneously reproduced by this kind of parametrisation.}
   {Given current constraints from asteroseismology, the core rotation rate of post-main sequence stars seems to be well explained by a process whose efficiency is regulated by the internal degree of differential rotation in radiative zones.}


   \maketitle
%
%-------------------------------------------------------------------

   \section{Introduction}
   \label{intro}
   The detection of splittings of mixed modes in post-main sequence stars enabled the measurement of the internal rotation rate of stars at different evolutionary phases, particularly in subgiants \citep{deheuvels14,deheuvels20}, red giant branch (RGB) stars in the hydrogen shell-burning phase \citep{beck12,deheuvels12,mosser12,dimauro16,dimauro18,gehan18}, red giants in the core-helium burning phase \citep{mosser12,deheuvels15,tayar19}, among others.
   This information combined with precise constraints on their structure and fundamental parameters such as the stellar mass and effective temperature give us information on how the physical processes redistributing angular momentum (AM) in stellar interiors must act through evolution.   
   In particular the first constraints on the internal rotation rate of red giants \citep{beck12} led to the conclusion that hydrodynamical processes usually adopted in stellar evolution computations such as meridional currents and shear instabilities \citep{zahn92} can not account for the core rotation rate of evolved low-mass stars \citep{eggenberger12, marques13, ceillier13}, pointing out to a missing physical process in stellar models.
   Different processes were proposed as candidates to explain the lack of AM transport in stellar interiors, such as internal magnetic fields \citep{spruit02,cantiello14,fuller19,takahashi21,eggenberger22}, wave-like perturbations such as internal gravity waves \citep{talon05,pincon17} or mixed modes \citep{belkacem15}, inward pumping of AM from convective envelopes \citep{kissin15}, among others.
   However, none of them gave a definite answer to the problem of the AM transport in stellar interiors so far.
   Particularly troublesome is the subgiant-red giant connection given the difficulty for a given transport process to reproduce the core rotation rate in both phases simultaneously \citep{eggenberger19c}. In the same way, the high efficiency of AM transport needed to provide a good overall agreement on the red-giant branch (RGB) can lead to some discrepancies with core rotation rates observed during the core-helium burning phase of intermediate-mass stars \citep{denhartogh20}.

   The increasing amount of results by extensive studies by the asteroseismology community on internal rotation \citep{mosser12,deheuvels14,deheuvels15,vanreeth16,gehan18,tayar19,deheuvels20,li20} and detailed studies on individual stars \citep{dimauro16,fellay21,salmon22,tayar22} enables us to explore new alternatives to the physical processes responsible for the internal AM transport at different evolutionary phases and different mass ranges.
   In a series of recent papers, the efficiency of the additional AM transport process that would be needed in stellar evolution models to fit the core rotation rate of evolved stars has been quantified \citep{denhartogh19,eggenberger19a,deheuvels20,moyano22} leading to the conclusion that the physical process needed should decrease its efficiency just after the end of the main sequence, and gradually become more efficient through early RGB and core-helium burning phase, as well as being more efficient in more massive stars in general.
   This kind of exploratory work enables us to gain information about the nature of the physical process across different stellar conditions.

   In a recent work \citet{spada16} showed that a physical process whose efficiency increases with the ratio of core-to-surface rotation rate can account for the internal rotation of RGB stars and is in mild agreement with that of subgiants.
   This kind of parametrisation for the AM transport efficiency was proposed to be related to the Azimuthal Magneto-Rotational Instability \citep[AMRI;][]{rudiger15}, a different version of the standard magneto-rotational instability \citep[MRI;][]{balbus91} in which the background magnetic field is mainly azimuthal and current-free (i.e. $B_{\phi} \propto 1/r$ with $r$, the radial distance) and the instability arises from non-axisymmetric modes as opposed to the MRI where the instability arises from axisymmetric perturbations in a medium with a poloidal field.
   The AMRI was explored in numerical simulations \citep{rudiger14,rudiger15,gellert16,guseva17} and its existence was verified in laboratory experiments \citep{seilmayer14}; this instability thus represents a possible candidate to the AM transport problem.
   \citet{spada16} argued that the transport efficiency of the AMRI increases as the contrast between the core and the surface increases.
   They thus parametrised the diffusion coefficient as a power law of the core-to-surface rotation rate and presented stellar evolution models of low-mass stars with a single mass value of $M=1.25 M_{\odot}$ in the subgiant and early red-giant phase finding a good agreement with the core rotation rate of RGB stars by fitting an exponent to the power law that is consistent with the scaling of the eddy viscosity of the AMRI.
   In this paper, we further explore this scenario for both low- and intermediate-mass stars until the end of the core-helium burning phase, and explore the role of the molecular viscosity.
   We benefit from a larger dataset of red giants in the hydrogen-burning shell phase with revised trends in the core rotation evolution \citep{gehan18}, as well as two early subgiants close the end of the main sequence \citep{deheuvels20}.

   In Sect. \ref{methods} we describe the physical ingredients of our models and the data used, in Sect. \ref{results} we present the models obtained and the comparison with the data and finally discuss its implications in Sect. \ref{discussion}.
   We conclude in Sect. \ref{conclusion}.

%--------------------------------------------------------------------
   \section{Physical ingredients}
   \subsection{Stellar evolution code and initial conditions}
\label{methods}
We compute stellar evolution models with the Geneva stellar evolution code \citep[{\fontfamily{qcr}\selectfont GENEC};][]{eggenberger08} taking into account the transport of angular momentum in radiative regions in the shellular rotation approximation \citep{zahn92}.
We include the advection of angular momentum by meridional circulation as well as diffusion by the shear instability. 
Convective zones are assumed to rotate rigidly.
The equation that describes the AM transport in radiative regions is given by
\begin{equation}
  \rho \frac{{\rm d}}{{\rm d}t} \left( r^{2}\Omega \right)_{M_r}
  =  \frac{1}{5r^{2}}\frac{\partial }{\partial r} \left(\rho r^{4}\Omega
  U(r)\right)
  + \frac{1}{r^{2}}\frac{\partial }{\partial r}\left(\rho D r^{4}
  \frac{\partial \Omega}{\partial r} \right) \
\label{eq_amt_genec}
\end{equation}
with $\Omega$ the horizontally averaged angular velocity, $\rho$ the density, $r$ the radial coordinate, $U$ the vertical component of the meridional circulation velocity and $D$ the total diffusion coefficient.
The coefficient $D$ takes into account the action of diffusive processes, which in our case are the shear instability ($D_{\rm shear}$) and the parametric diffusion coefficients that we explore ($D_{\rm add}$), so $D=D_{\rm shear} + D_{\rm add}$.
The solar metallicity is adopted for all the models with a solar chemical mixture as given by \citet{asplund09}, and the initial period is $P_{\rm ini}= 10$ days (see Sect 4.1 by \citet{moyano22} for a discussion on this choice).
We do not include braking by magnetic winds.
The rest of the parameters (e.g. overshooting, mass-loss, mixing-length parameter) where chosen in a similar way as in \citet{ekstrom12}.
The models are computed from the zero age main sequence (ZAMS) until the tip of the RGB for low-mass stars ($M \lesssim 2 M_{\odot}$), and until the end of the core-helium burning phase for models that do not go through the helium-flash ($M \gtrsim 2 M_{\odot}$).

In addition to {\fontfamily{qcr}\selectfont {GENEC}}, we use the {\fontfamily{qcr}\selectfont {MESA}} stellar evolution code version 15140  \citep{paxton11,paxton13,paxton15,paxton19} to compute low-mass ($M \lesssim 2 M_{\odot}$) models until the end of the core-helium burning phase \footnote{All the initial parameters and extensions necessary to reproduce our models are available at \url{https://zenodo.org/record/7305068}}.
We do this because {\fontfamily{qcr}\selectfont {GENEC}} is not prepared to routinely follow the angular momentum transport during the helium flash.
In {\fontfamily{qcr}\selectfont {MESA}} the AM transport is treated as a purely diffusive process \citep{paxton13} and its evolution is given by the equation
\begin{equation}
\left( \frac{\partial \Omega}{\partial t}\right)_m=\frac{1}{i}\left( \frac{\partial}{\partial m}\right)_t \left[ (4\pi r^2 \rho)^2iD \left( \frac{\partial \Omega}{\partial m}\right) \right]-\frac{2\Omega}{r}\left(\frac{\partial r}{\partial t}\right)_m\left( \frac{1}{2}\frac{\textrm{dln}i}{\textrm{dln}r}\right)
\label{eq_amt_mesa}
\end{equation}
where $i$ is the specific moment of inertia of a shell at a mass coordinate $m$, $D$ is the associated diffusion coefficient, and the rest of the variables keep their usual meaning.

We also include the additional diffusion coefficient that we explore ($D_{\rm add}$) in Eq. \ref{eq_amt_mesa} as part of the diffusion coefficient $D$ for our red-clump models computed with {\fontfamily{qcr}\selectfont {MESA}} on top of the meridional circulation in the Eddington-Sweet (ES) approach and the secular shear instability \citep[SSI; see][]{heger00}, thus $D=D_{\rm ES}+D_{\rm SSI}+D_{\rm add}$.
Solid-body rotation is also assumed in convective regions.
We recall that the main difference on the treatment of the AM transport between both stellar codes lies on the advective character of Eq. \ref{eq_amt_genec} which can lead to both inward and outward AM transport depending on the (non-trivial) sense of the circulation of each circulation cell \citep[e.g.][]{decressin09}.
  This can increase the shear rather than decreasing it as diffusive processes.
  However, for the initial masses and velocities that we adopt the meridional currents are not expected to act efficiently, thus their effects remain limited.

  Although different approaches are used we verified that a good agreement concerning the rotational evolution is obtained with both codes, especially in the upper RGB where the additional diffusion coefficient ($D_{\rm add}$) that we explore dominates.
%----------------------------------------------------------------------
\subsection{Angular momentum redistribution regulated by core-envelope coupling}
In line with previous efforts by \citet{spada16}, we explore the effect of a physical process whose efficiency scales with the contrast of rotation rate between the core and the surface.
We implement this as an additional diffusion coefficient in the equation of AM transport (Eq. \ref{eq_amt_genec}) following the prescription
\begin{equation}
  \label{eq_domegaratio}
  D_{\rm add}= D_{0} \left (\frac{\Omega_{\rm core}}{\Omega_{\rm surf}} \right )^{\alpha}
\end{equation}
where we take $\Omega_{\rm core}$ as the mean core rotation rate in the inner radiative layers, $\Omega_{\rm surf}$ is the surface rotation rate, and $D_{0}$ and $\alpha$ are free parameters.
In our models we take $\Omega_{\rm core}$ as the mean rotation rate in the region close to the core as sensed by gravity modes and whose expression is given by \citep{goupil13}
\begin{equation}
  \Omega_{\rm core}= \frac{\int_{0}^{r_{\rm g}}{\Omega \rm N_{\rm BV} dr/r}}{\int_{0}^{r_{\rm g}}{\rm N_{\rm BV} dr/r}}
\end{equation}
with N$_{\rm BV}$ the Brunt-V\"ais\"al\"a frequency, $r$ the radial coordinate, and $r_{\rm g}$ the upper boundary of the gravity-mode cavity.

 As mentioned in Sect. \ref{intro}, such a dependence on the core-envelope coupling was proposed to be related to the AMRI.
According to direct numerical simulations of Taylor-Couette flows, the eddy viscosity resulting from the onset of this instability can be fit with a relation that depends on adimensional numbers whose values are determined by the properties of the fluid.
In particular the turbulent eddy viscosity ($\nu_{\rm T}$) derived from direct numerical simulations was shown to scale as $\nu_{\rm T}/\nu \propto {Rm}/\sqrt{Pm}$ \citep{rudiger18} where $Pm=\nu/\eta$ is the magnetic Prandtl number, $Rm=Re Pm$ is the magnetic Reynolds number, with $\eta$ the magnetic diffusivity, and $\nu$ the molecular viscosity of the gas.
\citet{spada16} argue that the eddy viscosity in such experiments not only scales with $Rm$ and $Pm$ but also with the degree of shear between both cylinders and give the relation
\begin{equation}
  \label{nut_amri_s16}
  \frac{\nu_{\rm T}}{\nu} \propto \frac{Rm}{\sqrt{PmRa}} \left( \frac{\Omega_{\rm i}}{\Omega_{\rm o}} \right ) ^ {2}
\end{equation}
where $Ra$ is the Rayleigh number and the subindexes $i$ and $o$ refer to the values of the inner and outer cylinders in the Taylor-Couette simulations, respectively.
Taking into account only the dependence on the angular velocities, then one can approximate $\Omega_{\rm i} \sim \Omega_{\rm core}$ and $\Omega_{\rm o} \sim \Omega_{\rm surf}$, leading to the functional dependence proposed for the diffusion coefficient given by Eq. \ref{eq_domegaratio}.

Since ${Ra \equiv Q g (T_{o} - T_{i}) (R_{o} - R_{i})^{3}/\nu\chi}$ with $Q$ the coefficient of thermal expansion of the gas, $\chi$ the thermal conductivity, $g$ the local gravity, and $T_{\rm i,o}$ and $R_{\rm i,o}$ the radius and temperature of either the inner or outer boundaries, the right hand side of Eq. \ref{nut_amri_s16} does not depend on the molecular viscosity.
Then the turbulent viscosity alone is proportional to the molecular viscosity (i.e. $\nu_{\rm T} \propto \nu$).

    \begin{figure}
    \resizebox{\hsize}{!}{\includegraphics{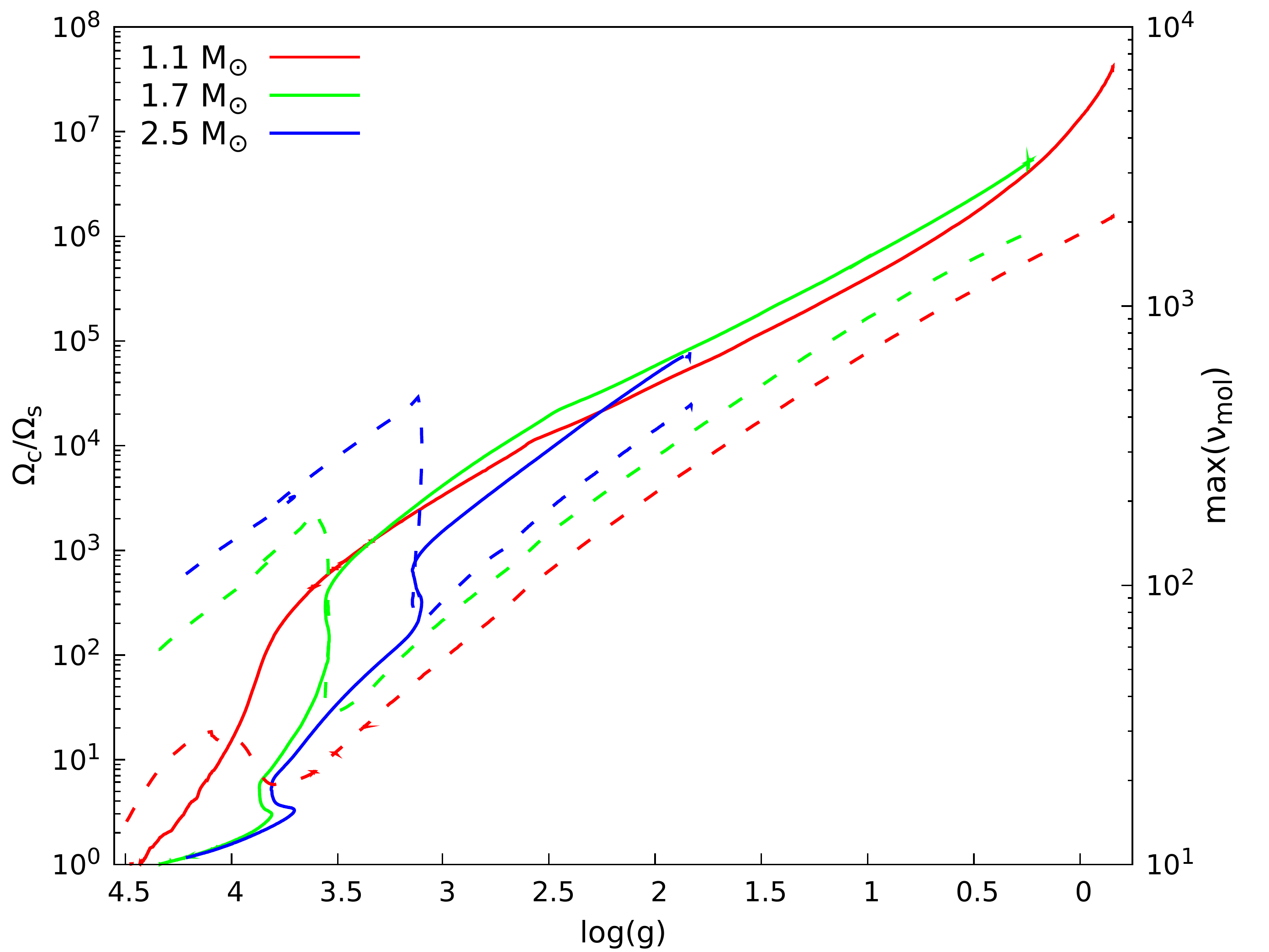}}
    \caption{Solid lines show the ratio of core-to-surface rotation rate while dashed lines show the maximum value of the molecular viscosity in the stellar interior through evolution from the zero age main sequence until the tip of the red giant branch, for different initial masses.      
      In this set of models the transport of angular momentum is driven only by hydrodynamical processes.}
    \label{visc_omegaratio}
    \end{figure}

Motivated by these findings we further explore the idea that the turbulent viscosity which regulates the transport of AM scales not only with the rotation contrast between core and surface but also with the intrinsic molecular viscosity of the gas.
This is a simple way to test the AMRI efficiency at different conditions in the stellar interior such as temperature and density, which occur in stars of different initial mass.
We consider the most extreme case in which the diffusion coefficient $D_{\rm add}$  is enhanced by the maximum value of the molecular viscosity in the radiative interior, thus we employ the following prescription
\begin{equation}
  \label{eq_visc_omegaratio}
D_{\rm add}=D_{1} \rm{max}(\nu_{\rm mol}) \left (\frac{\Omega_{\rm core}}{\Omega_{\rm surf}} \right )^{\alpha}
\end{equation}
where $D_{1}$ is a constant adimensional free parameter, and  $\nu_{\rm mol}$ is the kinematic molecular viscosity, whose expression for a mixture of hydrogen and helium is given by \citep{schatzman77}
\begin{equation}
  \nu_{\rm mol} = 2.2 \times 10^{-15} \frac{ T^{5/2}}{ \rho\ln \Lambda} \frac{1+7X}{8} \textrm{ [cm$^2$/s]}
\end{equation}
where $\ln\Lambda$ is the Coulomb logarithm and $X$ is the mass fraction of hydrogen.
Usually the maximum of the kinematic molecular viscosity is located in the radiative layers near the hydrogen burning shell.
The behaviour of both the rotation contrast between core and surface, and the maximum value of the molecular viscosity in the radiative interior for three representative models are shown in Fig. \ref{visc_omegaratio}.
  These models are computed taking into account only meridional circulation and shear instabilities.
  This shows that the molecular viscosity is expected to be higher in more massive stars, and grows as the star climbs the RGB.

%==========================================================================================
    \section{Stellar models}
    \label{results}
%------------------------------------------------------------------------------------------
\subsection{Red giant branch stars}
  \label{rgb_stars}
\begin{figure}
    \resizebox{\hsize}{!}{\includegraphics{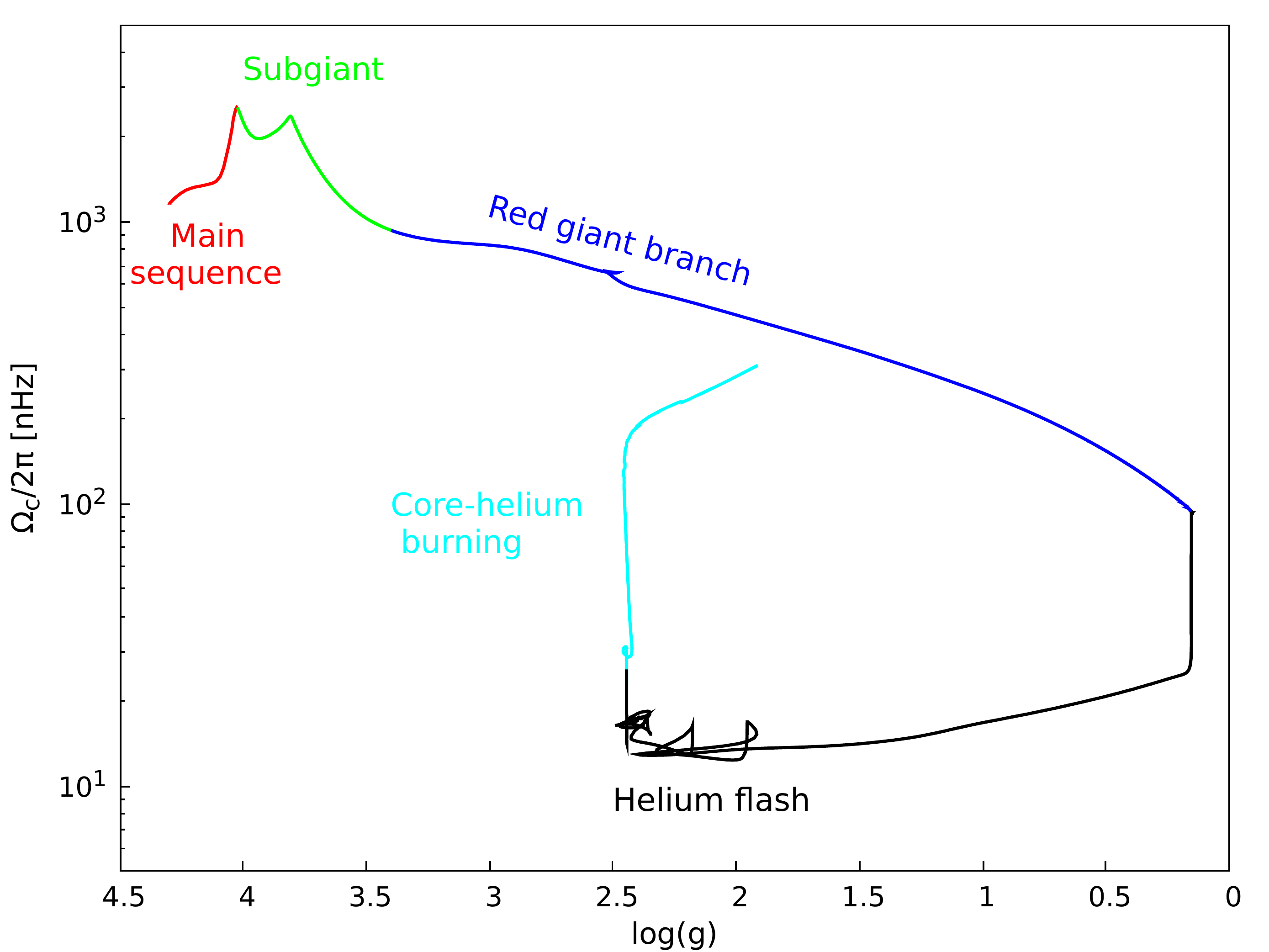}}
    \caption{Core rotation rate as a function of the surface gravity for a $1.3 M_{\odot}$ model with an initial period of $P=10$ days employing Eq. \ref{eq_domegaratio} with $D_{0}=50$ cm$^2$/s and $\alpha=2$, computed from the zero age main sequence until the end of the core-helium burning phase .
      The different evolutionary phases are indicated with different colours and labels.}
    \label{omegac_gsurf_cheb_labels}
    \end{figure}

\begin{figure}
    \resizebox{\hsize}{!}{\includegraphics{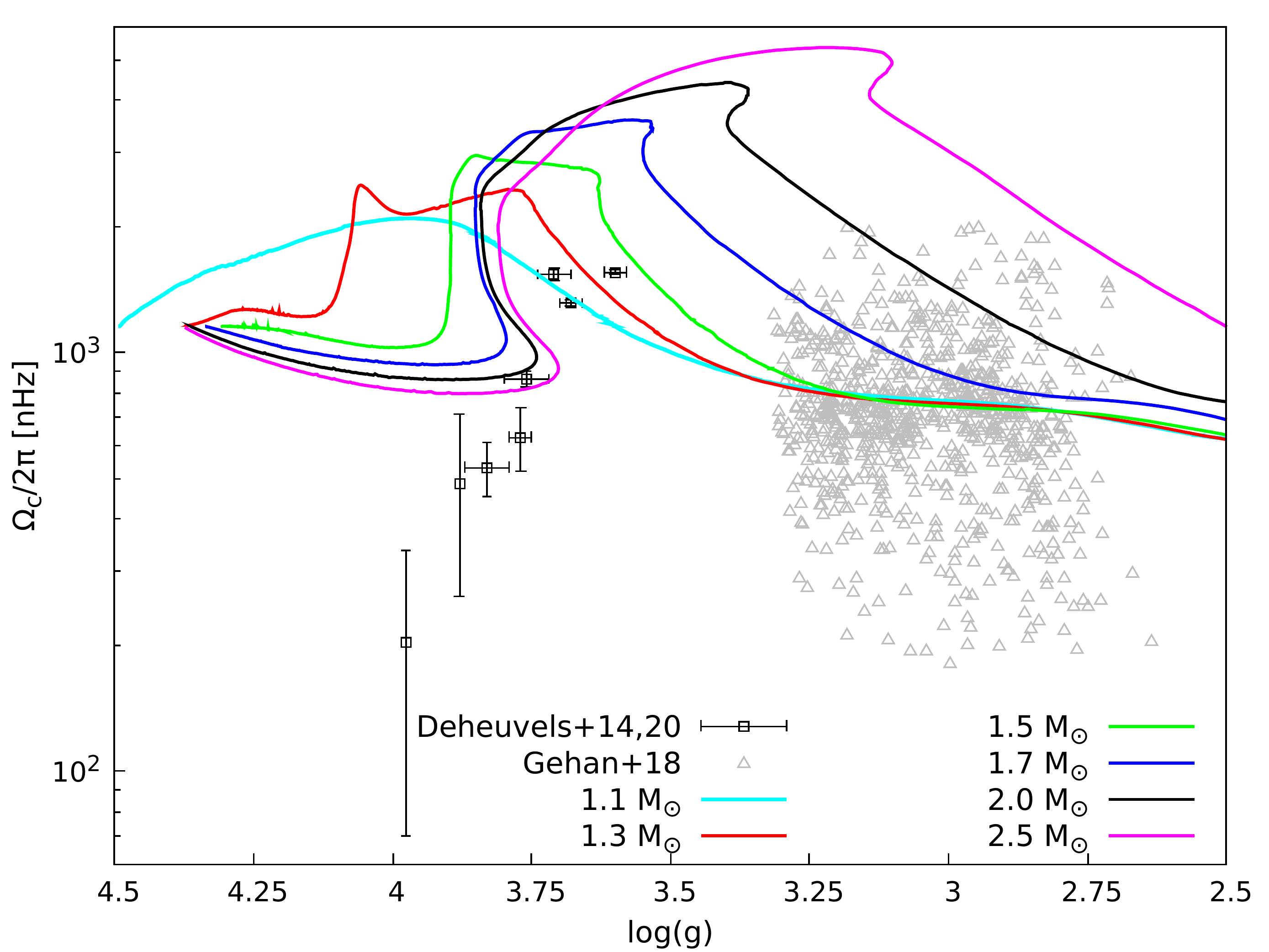}}
    \caption{Core rotation rate as a function of the surface gravity for models with different initial masses, as indicated in the figure.
      The models were computed using Eq. \ref{eq_domegaratio} with $D_{0}=50$ cm$^2$/s and $\alpha=2$.
      The black squares correspond to subgiants \citep{deheuvels14,deheuvels20} and the grey triangles to RGB stars \citep{gehan18}.
    }
    \label{omegac_gsurf_rgs}
    \end{figure}

  We computed rotating models with initial masses in the range $M=1-2.5 M_{\odot}$ from the ZAMS until the end of the core-helium burning phase using either the prescription given by Eq. \ref{eq_domegaratio} or \ref{eq_visc_omegaratio}, in addition to the shear instability and meridional circulation.
  An example of the evolutionary track in the $\Omega_{\rm core} - \log g$ diagram for a $1.3 M_{\odot}$ model computed from the ZAMS until the end of the core-helium burning phase is shown in Fig. \ref{omegac_gsurf_cheb_labels}.
    This model was computed with an initial period of $P=10$ days employing Eq. \ref{eq_domegaratio} with $D_{0}=50$ cm$^2$/s and $\alpha=2$.
  In Fig. \ref{omegac_gsurf_rgs}, we show the evolution of the core rotation rate for different initial masses using a diffusion coefficient given by Eq. \ref{eq_domegaratio} and a calibrated value of $D_{0}= 50$ cm$^2$/s and $\alpha=2$.
  The models with initial masses of $M=1.1, 1.3$ and $1.5 M_{\odot}$ can reproduce simultaneously the core rotation rate of the three subgiants with the fastest spinning cores around $\log g \sim 3.7$, and the core rotation rate of the bulk of RGB stars, which is in the range $\Omega_{\rm c}/2\pi \sim 600 - 800$ nHz.

    We chose a value of $\alpha=2$ to fit the mean flat trend in the evolution of the core rotation rate for RGB stars \citep{gehan18}.
    While $D_{0}$ regulates the maximum core rotation rate that can be achieved during the subgiant and early red-giant phase, $\alpha$ regulates the speed with which the core is decelerated and thus sets the slope of the rotational evolution in the RGB phase; the higher the value of $\alpha$, the steeper the slope of the core rotation rate during the RGB.
  Thus the large sample of RGB stars given by \citet{gehan18} constrains this value to $\alpha \sim 2$ for low-mass stars.
  These are not completely new results since they were already presented by \citet{spada16} for a $1.25 M_{\odot}$ model, but we confirm in this work that this scenario is valid in the mass range $M \sim 1 - 1.5 M_{\odot}$.  
  However, for more massive stars ($M \gtrsim 1.7 M_{\odot}$), the core rotation rate decreases too steeply and thus is not able to reproduce the apparently flat trend seen for RGB stars.
  This occurs in part because the radius of these stars is larger on the RGB compared to lower-mass stars and hence the surface rotation rate is lower, which increases the value of the diffusion coefficient and hence the deceleration rate of the core.
  %
  %  

%----------------------------------------------------------------------
%----------------------------------------------------------------------
  \subsection{Low-mass core helium burning stars}
\begin{figure}
    \resizebox{\hsize}{!}{\includegraphics{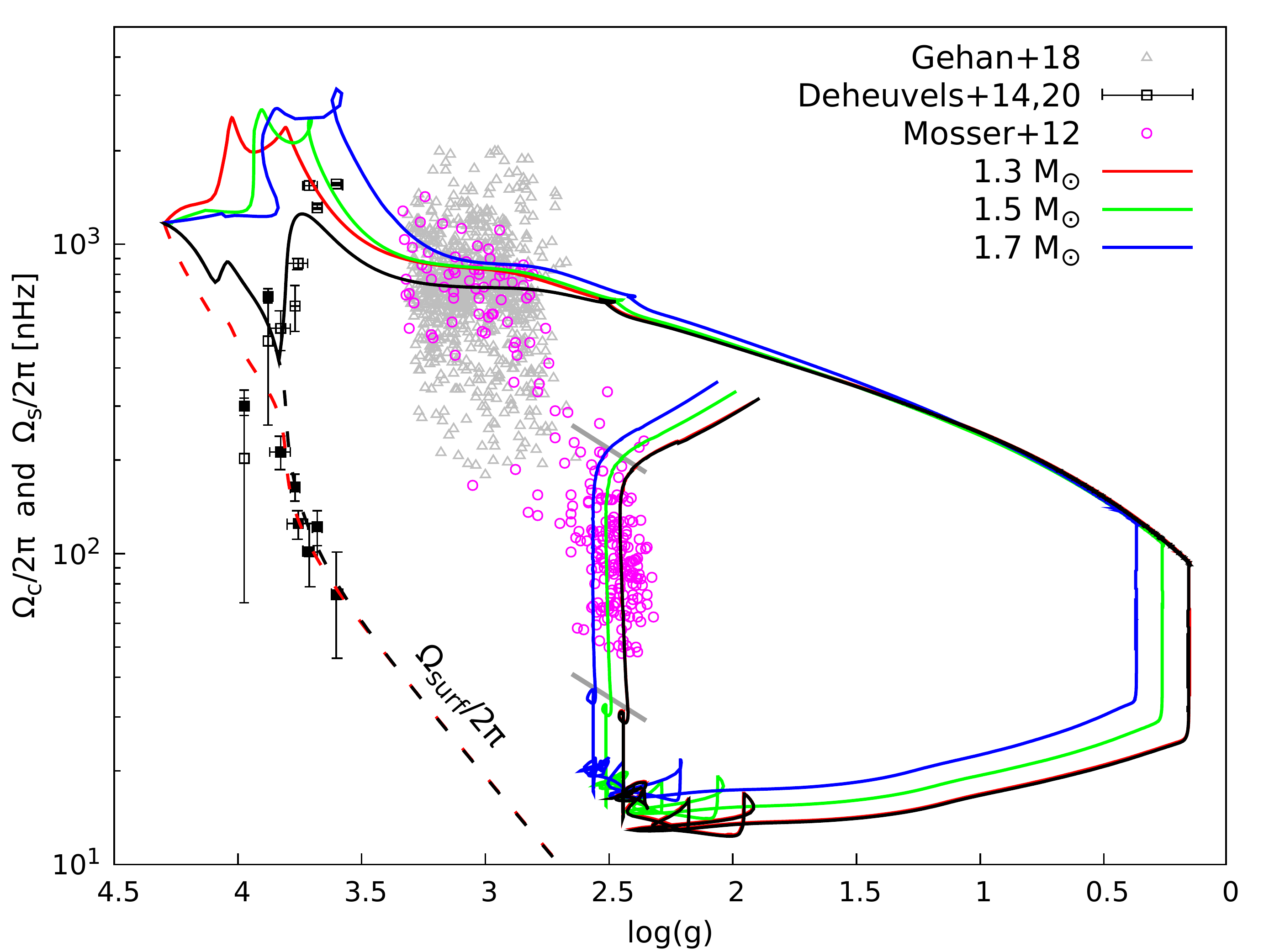}}
    \caption{Evolution of the core rotation rate for models with an additional diffusion coefficient of the form $D_{\rm add}=D_{0} (\Omega_{\rm core}/\Omega_{\rm surf})^2$, with $D_{0}= 50\ \rm{cm^2/s}$.
      The black filled(empty) symbols show the surface(core) rotation rate of the eight subgiants presented by \citet{deheuvels14,deheuvels20}.
      The grey symbols show the core rotation rate of stars in the lower RGB \citep{gehan18}.
      The magenta circles correspond to red giants presented by \citet{mosser12} with those grouped around $\log g \sim 2.5$ in the core-helium burning phase (we only include those with an estimated mass $M \lesssim 2 M_{\odot}$).
      The red and black dashed lines show the surface rotation rate of the $1.3 M_{\odot}$ model.
      The black-line shows a 1.3 M$_{\odot}$ model for which we enforced rigid rotation until the location of the second subgiant at $\log g \sim 3.9$.
      The grey lines at $\log g \sim 2.5$ mark the region where the central abundance of helium is in the range $Y=0.9 - 0.1$, i.e. the long and stable core-helium burning phase.
    }
    \label{omegac_gsurf_redclump}
  \end{figure}

  We computed models until the end of the core-helium burning phase for our low-mass models with $M= 1.3, 1.5$ and $1.7 M_{\odot}$ for which constraints on the core rotation rate are available \citep{mosser12}.
  In Fig. \ref{omegac_gsurf_redclump}, we show the evolution of the core rotation rate with an additional diffusion coefficient following Eq. \ref{eq_domegaratio}, with $D_{0}=50$ cm$^2$/s and $\alpha=2$, from the ZAMS until the end of the core helium burning phase.
  This is a continuation of the evolutionary scenario presented in Fig. \ref{omegac_gsurf_rgs} for RGB stars since we adopt the same diffusion coefficient.
  Our models with $M=1.3, 1.5$ and $1.7 M_{\odot}$ can reproduce the core rotation rate of the three fastest subgiants as well as their surface rotation rates, shown by the dotted line (only for the $1.3 M_{\odot}$ model).
  They can also reproduce the apparently flat trend of red giants in the hydrogen-shell burning phase (at $\log g \sim 3.3 - 2.9$) and the core rotation rate of core-helium burning stars for different masses (see red, green and blue lines in Fig. \ref{omegac_gsurf_redclump}).
  The models spend $\sim 85 \%$ of their core-helium burning time with their core spinning at $\Omega_{\rm core}/2\pi \sim 40 - 200$ nHz (denoted by grey lines), in agreement with the asteroseismic constraints given by \citet{mosser12} shown as magenta circles at $\log g \sim 2.5$.  
  This is achieved without changing any parameter during the evolution until the end of the core-helium burning phase.
  We also verify, as proposed by \citet{spada16}, that if one enforces solid body rotation until a given time during the subgiant branch, one can reproduce the increase of the core rotation rate seen for the subgiants.
  We show this for a $1.3 M_{\odot}$ model for which we enforced solid body rotation until roughly the location of the second subgiant seen in Fig. \ref{omegac_gsurf_redclump} whose internal rotation is consistent with that of a solid body \citep{deheuvels20}.
  Afterwards, the core rotation rate converges towards the RGB bump, reproducing in a similar way the constraints for RGB stars and red clump stars.

  In this series of models, the core rotation rate decreases by roughly one order of magnitude from the RGB base until the tip.
  This happens because, as the star ascends the RGB, the core contracts and the envelope expands until the tip, which in turn increases the rotation contrast between core and surface and thus increases the value of the diffusion coefficient, slowing down the core progressively.
  This is even more pronounced if one chooses higher values of $\alpha$ which lead to lower core rotation rates towards the RGB tip.  
  Once the models reach the tip, the helium is ignited off-center in degenerate conditions, leading to the helium flash, which produces enough energy to expand the central layers and thus allows the core to expand and slow down due to its increased moment of inertia.
  This is why the core rotation rate decreases abruptly at $\log g \sim 0.5 - 0.2$ for the three models presented in Fig. \ref{omegac_gsurf_redclump}.
  This behaviour is general of any rotating model and does not depend on the physical process adopted, because the timescale in this phase is too short ($\sim 2-3$ Myr) to allow for AM transport.
  Afterwards, the star contracts, increasing its surface gravity, and goes through a series of helium sub-flashes, which are seen as small loops around $\log g \sim 2 - 2.5$, until the core becomes non-degenerate and the stable core-helium burning phase can begin.
  As mentioned before, once the star settles into the stable core-helium burning phase, it spends $\sim 85$ \% of its core-helium burning time with core rotation rates $\Omega_{\rm core}/2\pi \sim 40 -200$ nHz in agreement with asteroseismic constraints.
      Towards the end of the core-helium burning phase, the core shifts to He- and H-shell burning and the subsequent CO core contracts while the envelope expands, leading to a spin-up of the core and a decrease in surface gravity, which occurs at $\log g \sim 2.5$ and $\Omega_{\rm c}/2\pi \sim 200$ nHz.
  
  In these models, the core is found to spin-up during the core-helium burning phase.
  This behaviour seems a priori to be counter-intuitive, since under local conservation arguments, the core should spin at a rather constant rate since its size (in radius) does not change enough to alter significantly its moment of inertia.
  Furthermore, if the core rotates faster than the surface and we assume that the angular velocity is distributed smoothly, then the diffusion of AM from the core to the outer layers should spin down the core.
  However, this counter-intuitive behaviour occurs because of the particularly low core rotation rate at the RGB tip and the dynamics of the He-flash.
  It can be explained by analysing the rotation profile, which in turn is essentially explained by the contraction and expansion (hence increase or decrease of the density) of the layers since the timescales are too short for AM transport to take place \footnote{See \citet{iben13} for a detailed discussion}.

\begin{figure}
    \resizebox{\hsize}{!}{\includegraphics{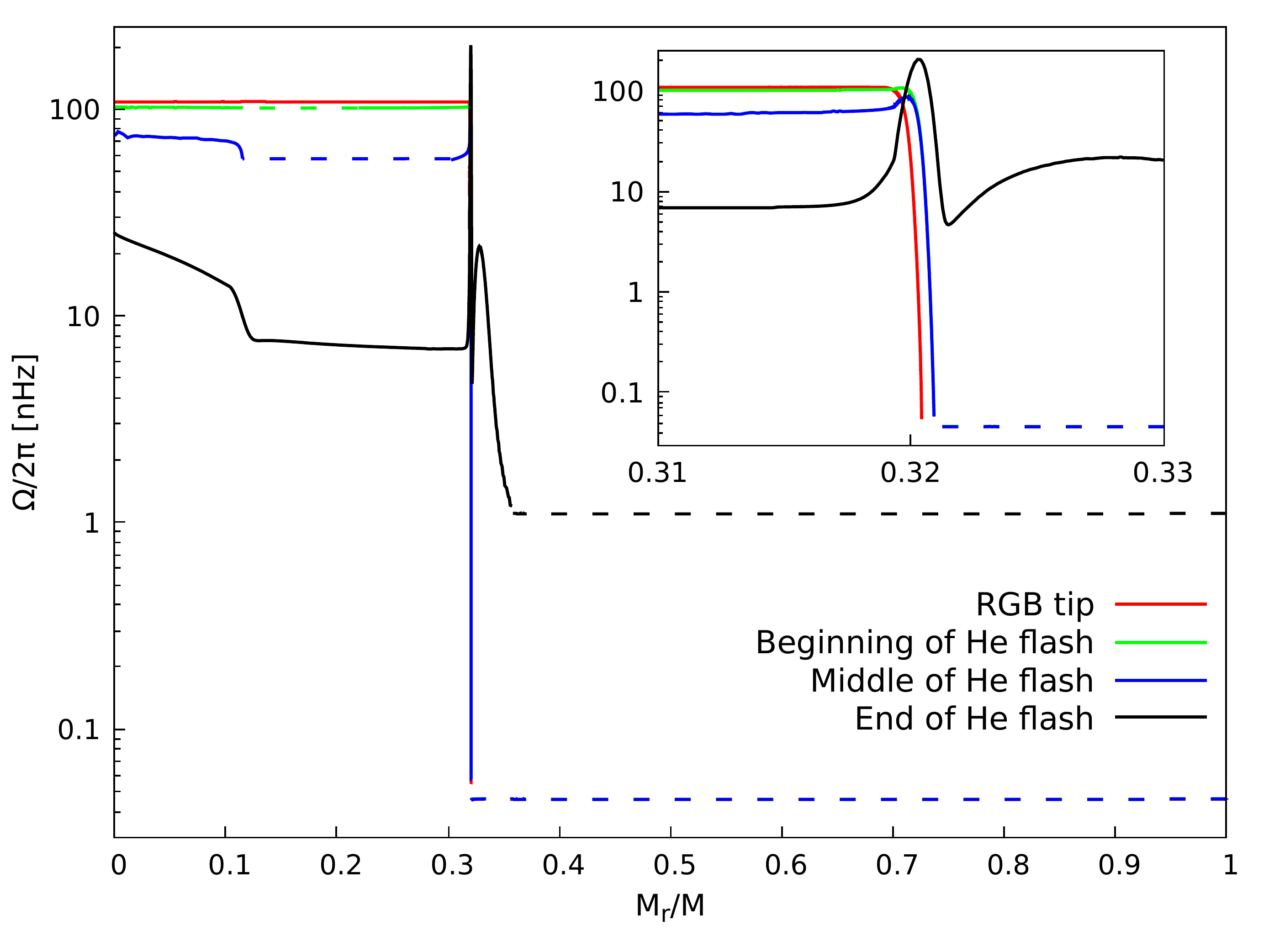}}
    \caption{Angular velocity as a function of the mass coordinate at different moments during the first helium flash of the $1.5 M_{\odot}$ model presented in Fig. \ref{omegac_gsurf_redclump}.
      Radiative regions are shown by solid lines while dotted lines show convective regions.
      The inset around $M_{\rm r}/M \sim 0.32$ shows a zoom into the location of the H-burning shell.
    }
    \label{rotprof_heflash}
  \end{figure}

    The evolution of the rotation profile during the first He-flash is shown in Fig. \ref{rotprof_heflash}.
    During the first He-flash, the peak of nuclear energy is located at the base of the convective He-shell, around $M_{\rm r}/M \sim 0.12$ in Fig. \ref{rotprof_heflash}, and most of the energy released produces an expansion of the inner layers, but causes the layers at the base of this convective shell to spin down faster than the surrounding layers.
    %, but spins down faster the layers at the base of this convective shell.
  This explains that, in Fig. \ref{rotprof_heflash}, the model at roughly the middle of the He-flash exhibits a rotation profile with a dip close to the base of the helium convective shell, where the uniform rotation rate is due to convection (marked by dotted lines) and the rotation rate is higher in the neighbouring hydrogen-rich layers located at $M_{\rm r}/M \sim 0.32 (M_{\rm r} \sim 0.465 M_{\odot})$.
    As the core expands and the regions above move outwards, the hydrogen-burning shell cools down and its nuclear energy generation rate decreases, forcing the star to contract and to release gravitational energy, decreasing the luminosity on its way to the horizontal branch.
    This contraction occurs outside the He-convective shell and across the whole convective envelope, which leads to faster rotating layers near the hydrogen-rich layers and leads to the rotation profile shown by the black line in Fig. \ref{rotprof_heflash}.
    %See figure 17.1.17 from Iben+2013
    At this stage, the star has already contracted, which can be deduced from its much higher surface rotation rate.
    During the subsequent helium sub-flashes, a similar scenario occurs regarding the rotation rate of the layers above the convective helium shell.
  These layers remain rotating faster than the core until the beginning of the core-helium burning phase.
  And given the high degree of shear in those rapidly rotating layers, AM is easily diffused to the core by e.g. the shear instability or even viscous effects, which spins it up during the long and stable core-helium burning phase.
  This is the reason why the core rotation rate increases during the core-helium burning phase in these models.
  %
  %----------------------------------------------------------------------
  %SECONDARY CLUMP
  %----------------------------------------------------------------------
  \subsection{Intermediate-mass core helium burning stars}
\begin{figure}
    \resizebox{\hsize}{!}{\includegraphics{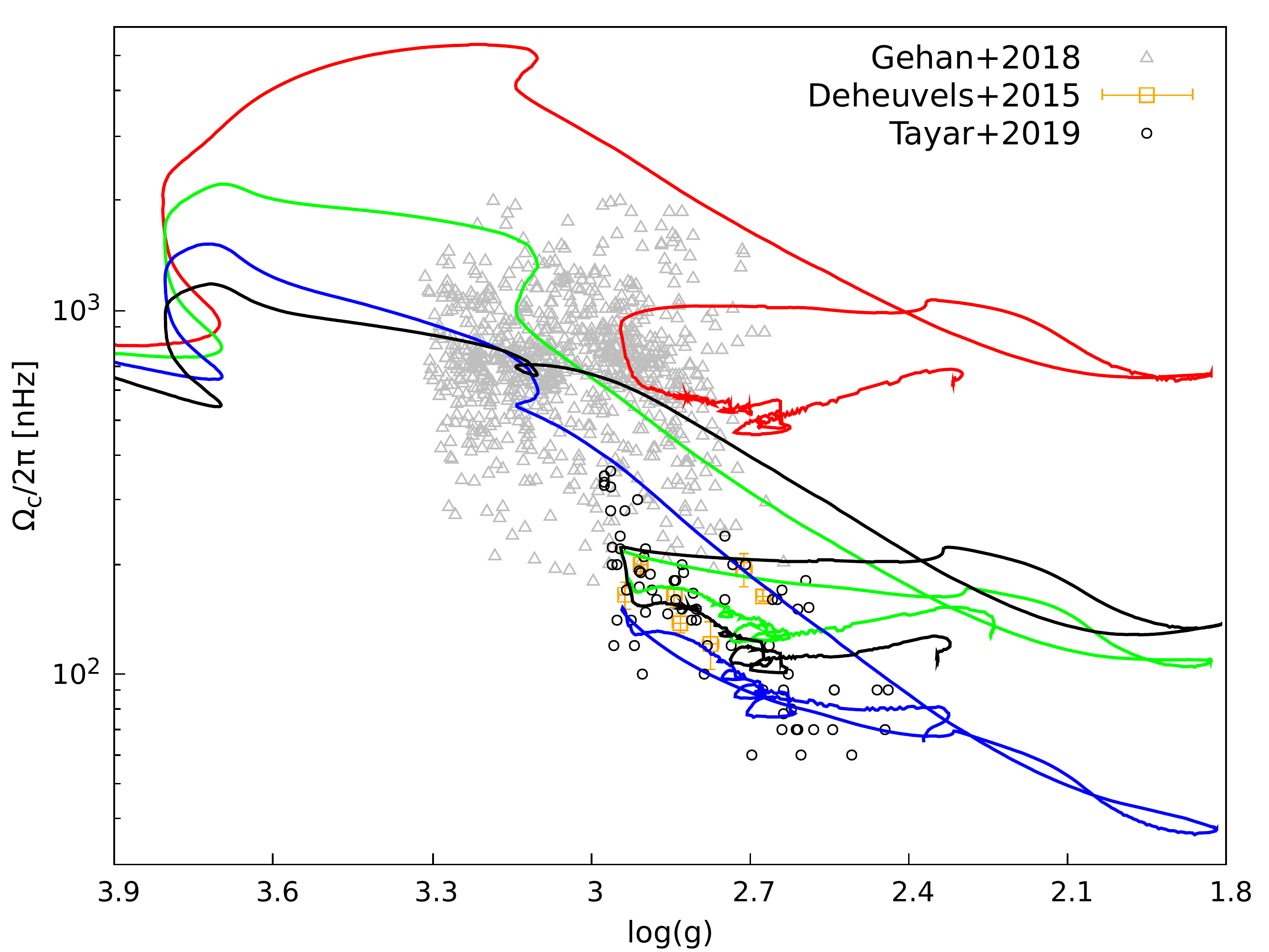}}
    \caption{Core rotation rate as a function of the surface gravity for $2.5 M_{\odot}$ models until the end of the core-helium burning phase.
      The models are computed with Eq. \ref{eq_domegaratio} with $\alpha=2$ and $D_{0}= 50$ and 2000 cm$^2$/s for the red and green lines, respectively.
      The blue and black-line models are computed with Eq. \ref{eq_visc_omegaratio} (including the enhancement by the molecular viscosity) using $\alpha=2$, $D_{1}= 50$ and $\alpha=1$, $D_{1}=300$, respectively.
      The grey triangles correspond to red giants in the lower red giant branch \citep{gehan18} and the black circles and orange squares to intermediate-mass core-helium burning stars in the secondary clump \citep{deheuvels15,tayar19}.
    }
    \label{omegac_gsurf_2ndclump}
  \end{figure}

  We also follow the evolution of our intermediate-mass models of $2.5 M_{\odot}$ until the end of the core-helium burning phase with different prescriptions for the transport of AM (see Fig. \ref{omegac_gsurf_2ndclump}).
    Most of the stable core-helium burning phase ($0.1 \lesssim Y \lesssim 0.9$) occurs once the star already descended from the RGB tip (located at $\log g \sim 1.8$) and starts expanding as it becomes more luminous.
  This spans a range of surface gravities $\log g \sim 2.9 - 2.6$ in our models.
  In all three models, the core spins down during this phase.

  In Fig. \ref{omegac_gsurf_2ndclump}, we compare our different models with the data from secondary-clump stars from \citet{deheuvels15} and \citet{tayar19}.
  The red line model corresponds to our $2.5 M_{\odot}$ model presented in Fig. \ref{omegac_gsurf_rgs} computed with Eq. \ref{eq_domegaratio} using $\alpha=2$ and $D_{0}=50$ cm$^2$/s extended until the end of the core-helium burning phase.
  It does not reproduce the core rotation rate of RGB stars and we show here that it does not either reproduce the constraints for stars in the core-helium burning phase.
  If we re-calibrate the constant variable to $D_{0}=2000$ cm$^2$/s (without changing the power $\alpha$) to reproduce the RGB stars then we can automatically reproduce the rotation rate of secondary-clump stars; this is shown by the green-line model.
  As mentioned in Sect. \ref{rgb_stars}, the core rotation of our most massive models is not in agreement with the data of RGB stars.
  Regarding this point, we note that the molecular kinematic viscosity is higher in general for more massive stars, mainly because of higher internal temperatures, and increases monotonically through evolution on the RGB (see Fig. \ref{visc_omegaratio}).
  To test whether the AM transport efficiency including an enhancement by the molecular viscosity can bring the models into agreement with the data for the massive stars in our sample, we computed models following Eq. \ref{eq_visc_omegaratio}.
  When we include this additional factor and consider $\alpha=2$ and $D_{1}=50$ the core rotation rate of the $2.5 M_{\odot}$ model close to the RGB base is $\Omega_{\rm core}/2\pi \sim 600$ nHz in agreement with the data of RGB stars, but decreases too steeply with evolution, in contradiction with the apparently flat trend (see blue line in Fig. \ref{omegac_gsurf_2ndclump}).
  Interestingly, once we take this enhancement into account, the core rotation rate of secondary clump stars can be reproduced.
  We find that a good fit able to reproduce both the apparent flat trend in core rotation rate of RGB stars and the core rotation rate of secondary clump stars can be obtained if one includes the enhancement by the molecular viscosity as in Eq. \ref{eq_visc_omegaratio} with $\alpha=1$ and $D_{1}= 300$, shown by the black-line model in Fig. \ref{omegac_gsurf_2ndclump}. 
    This provides evidence that if the AM transport efficiency is regulated by the core-envelope coupling, once the core rotation rate of RGB stars is reproduced then secondary clump stars can also be well reproduced.  
    A value of $\alpha=1$ is different from the values of $\alpha \simeq 2 - 3 $ previously suggested by \citet{spada16}.
    This could point to a mass-dependence for the efficiency on the AM transport by this parameterisation; the efficiency of AM transport may have a weaker dependence on the degree of differential rotation for higher-mass stars.
  %
  % COMPARISON WITH ADDITIONAL VISCOSITIES
  % ----------------------------------------------
  \subsection{Comparison with additional viscosities}
\begin{figure}
    \resizebox{\hsize}{!}{\includegraphics{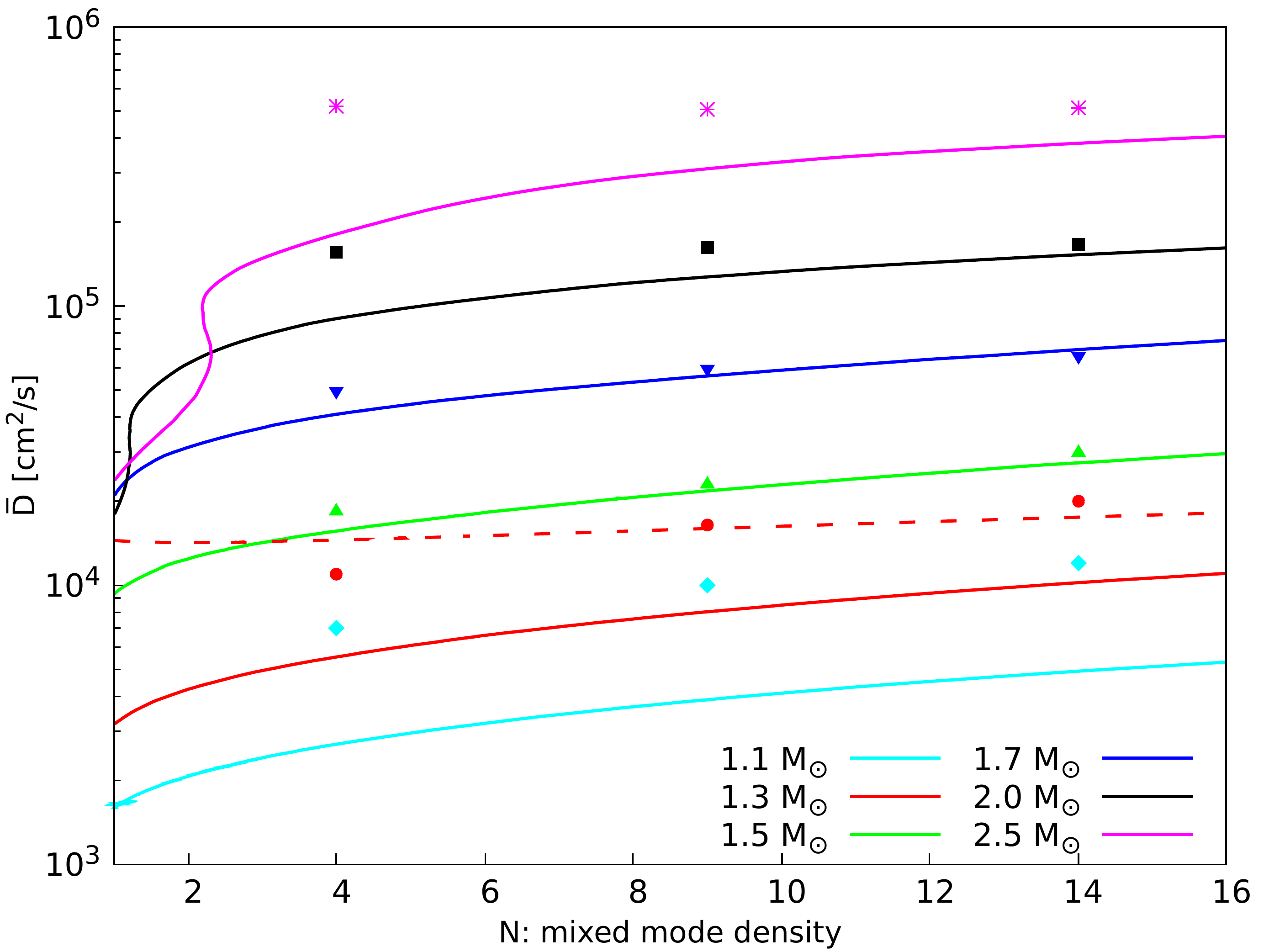}}
    \caption{Time-averaged diffusion coefficient as a function of the mixed mode density for different masses computed with Eq. \ref{eq_domegaratio} using $\alpha=2$ and $D_{0}=50$ cm$^2$/s.
      The points correspond to the additional viscosities estimated by \citet{moyano22} to reproduce the core-rotation rate of RGB stars at different mixed-mode densities.
      The red-dashed line corresponds to a model where a constant value of $D=2 \times 10^4$ cm$^2$/s was used until $\log g \simeq 3.9$.
    }
    \label{dcoeff_mixmod}
  \end{figure}

  In previous works, we estimated the efficiency of the additional AM transport process needed to reproduce the core rotation rate of red giants via a parametric approach \citep{moyano22}, in particular to reproduce the apparently flat trend of core rotation rates of RGB stars.
  In addition to the direct comparison between predicted and observed core rotation rates presented in Fig. \ref{omegac_gsurf_rgs}, we can then also compare the values of the diffusion coefficients predicted by Eq. \ref{eq_domegaratio} with the mean values along this phase determined from our previous estimates to compare its behaviour with mass and evolution.
  Since the diffusion coefficient (Eq. \ref{eq_domegaratio}) employed in the present models changes through the evolution, while the mean additional viscosities presented by \citet{moyano22} are defined as being constant, a direct comparison between both quantities would not be accurate, since the former are instantaneous values at a given time and the latter would represent mean values through the evolution until a given constraint is satisfied.
  To reconciliate these two aspects and make a fairer comparison between both approaches, we take the time average value of the diffusion coefficient $D_{\rm add}$ as follows
  \begin{equation}
    \bar{D}(t)= \frac{1}{t-t_{\rm TAMS}} \int_{t_{\rm TAMS}}^{t} D(t') dt'
  \end{equation}
  where $t$ is the current age and $t_{\rm TAMS}$ the age at the terminal age main sequence (TAMS).
  This approach can give us an estimate of the additional viscosity that would have been needed through the evolution until a certain time $t$ to obtain the core rotation rate of a given model.
  The comparison is shown in Fig. \ref{dcoeff_mixmod} for the models in the whole mass range studied as a function of the mixed mode density.
  The mixed mode density, defined as $\mathcal{N}= \Delta\nu /(\Delta\Pi_{1}\nu_{\rm max}^2)$ with $\Delta\nu$ the large frequency separation, $\nu_{\rm max}$ the frequency of the maximum oscillation signal, and $\Delta\Pi_{1}$ the asymptotic period spacing of dipole modes, was proposed as an asteroseismic indicator of the evolution through the lower RGB with the advantage of showing no strong correlation with the stellar mass \citep{gehan18}, allowing us to study the evolutionary trends for different masses.
  The absolute values are within the ranges expected, where the disagreements for the 1.1 and 1.3 $M_{\odot}$ models may arise because in Fig. \ref{omegac_gsurf_rgs} the models with masses $M= 1.1 - 1.3 M_{\odot}$ evolve at core rotation rates slightly higher than $\Omega_{\rm core}/2\pi \sim 700$ nHz whereas \citet{moyano22} estimated the additional viscosity needed for the models to achieve that core rotation rate through the range of mixed mode densities presented.
  Interestingly, the trends with mass are roughly satisfied with the parametrisation given by Eq. \ref{eq_domegaratio}; i.e. the efficiency increases roughly two orders of magnitude in this mass range for early RGB stars.
  
  \citet{moyano22} also studied how the efficiency should increase as evolution proceeds to keep the core rotating at the expected rate ($\Omega_{\rm core}/2\pi \sim 700$ nHz) during the early RGB and found that the additional viscosity for low-mass stars should increase by roughly a factor two.
  Here we find that the behaviour of the normalized values is similar for the low-mass stars, i.e. the time-averaged diffusion coefficient increases by roughly a factor two in the early RGB in the range of mixed mode densities $\mathcal{N} \sim 4 - 14$.
  However, in the case of more massive stars, the increase is much higher than expected, which can be directly understood from Fig. \ref{omegac_gsurf_rgs}, where the values of the core rotation rate of the massive models ($M= 2.0, 2.5 M_{\odot}$) are found to decrease during the RGB, in contradiction with the data.
  To take the time average of the additional diffusion coefficients we choose as a starting point the TAMS, since the evolution of the core rotation is dominated by what happens during the subgiant phase (see e.g. \citet{eggenberger17, eggenberger19a}).
  However if one enforces solid body rotation during the main sequence (as was done by \citet{spada16}), the time-averaged diffusion coefficients would have lower values because the core rotation rate in this case reaches lower values at the TAMS and thus remains lower during the subgiant and early RGB until it converges during the RGB.  
  The effect mentioned on the core rotation rate can be seen comparing the black and red lines in Fig. \ref{omegac_gsurf_redclump}, where we enforced solid-body rotation (black line) until the location of the second subgiant of the sample of \citet{deheuvels20} in the $1.3 M_{\odot}$ model.
  In this case once the models reach the RGB bump (seen at $\log g \sim 2.5$) both models converge to the same core rotation rate and evolve identically, erasing any previous rotational history.
  However the behaviour for the normalized values remains similar, the time-averaged diffusion coefficient increases by roughly a factor two.
   We also note that considering a constant diffusion coefficient of $D=2 \times 10^4$ cm$^2$/s from the TAMS until the location of the second subgiant would make our average diffusion coefficient slightly larger and would then be in agreement with the values inferred for red giants.
  This is shown by the red-dashed line in Fig. \ref{dcoeff_mixmod}.  
  This diffusion coefficient used until the location of the second subgiant predicts a slightly decoupled envelope, with a contrast of core-to-surface rotation rate of $\Omega_{\rm c}/\Omega_{\rm s} \simeq 1.4$.
  %NOTE: AN APPROPIATE VALUE TO KEEP RIGID THE 1.3 MSUN MODEL IS 1D6
  
  %================================================================================  
  %================================================================================  
  \section{Discussion}
  \label{discussion}
  Considering only hydrodynamical processes in stellar evolution computations, the rotation contrast between the core and the surface increases with evolution as a result of the surface expansion and the contraction of the core.
  Assuming that the rotation profile is smooth (i.e. it does not have sharp discontinuities), the core-envelope coupling is a global measure of the shear in the radiative regions, where we expect instabilities to redistribute AM.
  In previous works \citep{eggenberger17,moyano22}, we showed that the efficiency of the AM transport should increase both with mass and evolution along the RGB to satisfy the constraints from asteroseismology.
  Since the properties mentioned above go in this direction, we revisited the computations presented by \citet{spada16}, which they argue could be related with the AMRI.
    
  \citet{spada16} showed that employing a diffusion coefficient that scales with the inverse of the core-envelope coupling (see Eq. \ref{eq_domegaratio}) and enforcing rigid rotation until approximately the middle of the subgiant phase would reproduce simultaneously the core rotation rate of subgiants and red giants (see the upper panel of their Fig. 4).
  Adopting the same approach and following the evolution until the end of the core-helium burning, we find that all three evolutionary phases can be well reproduced.
  Moreover, recent results by \citet{deheuvels20} show that two young subgiants close to the TAMS are rotating roughly as solid bodies, consistent with the hypothesis above mentioned.
  Regarding the main sequence, there is some evidence from $\gamma$ Dor stars that the core is strongly coupled to the envelope, with a core that spins not faster than $\sim 5$ \% than its surface \citep{li20,saio21}, while models with only hydrodynamical processes attain a contrast of $\Omega_{\rm core}/\Omega_{\rm surf} \sim 2$ by the middle of the MS.
  Also, recent theoretical works on AM transport in $\gamma$ Dor stars combining AM constraints on both surface and core, favor highly efficient transport of AM during the MS, such that solid-body rotation in radiative zones is a good approximation (Moyano et al., in prep.).
  This group of stars represents the MS counterpart of the subgiants and red giants studied in this work since their mass range is $M \sim 1.4 - 2.0 M_{\odot}$.
  However, for our low-mass models, the core rotation rate during the RGB onwards is roughly independent on the rotational history during the main sequence.

  \begin{figure}[h!]
    \resizebox{\hsize}{!}{\includegraphics{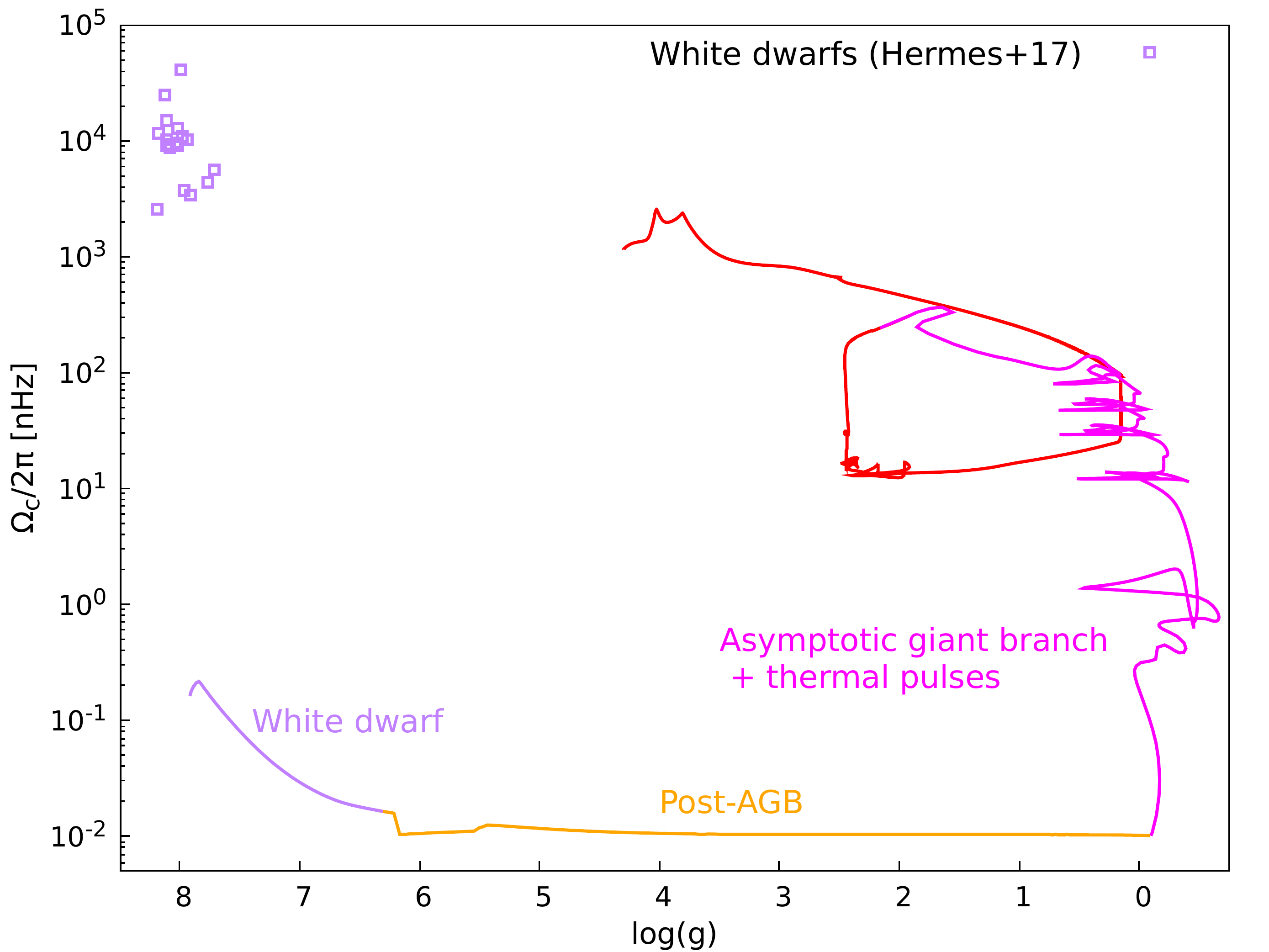}}
    \caption{Core rotation rate as a function of the surface gravity for a $1.3 M_{\odot}$ model with an initial period of $P=10$ days.
      Angular momentum transport is driven by meridional circulation, shear instabilities, and the parametric diffusion coefficient of the AMRI (Eq. \ref{eq_domegaratio}).
      Evolutionary phases are indicated in the figure.
      The data points correspond to the rotation periods of white dwarfs \citep{hermes17}.
    }
    \label{omegac_gsurf_wd}
    \end{figure}

  We note that the diffusion coefficient explored in this work can not couple efficiently the core with the envelope during the main sequence, so that a more efficient process, possibly scaling with other properties of the star \citep[as for instance transport by the Tayler instability, see][]{eggenberger22b}, would be needed during the MS and early subgiant branch.
 Concerning later evolutionary phases, and in particular evolution along the asymptotic giant branch, we found that the envelope expands and leads to very low surface rotation rates, which in turn leads to a very high rotation contrast between core and surface.
 Because of this, the diffusion coefficient we adopted reaches very high values ($\log D_{\rm add} \sim 8 - 10$ cm$^2$/s) which slows-down the core to periods in strong disagreement with those of white dwarfs.
 % NEW TEXT 2/11
 Specifically, our models with initial masses of 1.3, 1.5 and 1.7 $M_{\odot}$ presented in Fig. \ref{omegac_gsurf_redclump} reach core rotation periods of $P_{\rm core} \sim 30 - 300$ days while those of observed white dwarfs in the corresponding mass-range ($M \sim 0.5 - 0.6 M_{\odot}$) are of $P_{\rm core} \sim 10 -100$ hours \citep{hermes17}.
 This along with the whole evolutionary path towards the white dwarf phase is shown in Fig. \ref{omegac_gsurf_wd}.
 For clarity, we show only the evolutionary track of a single mass and we do not show the data of subgiants nor red giants.
   The evolution for our two models of 1.5 and 1.7 $M_{\odot}$ is similar.
  The evolution until the end of the core-helium burning phase corresponds to the same $1.3 M_{\odot}$ model presented in Fig. \ref{omegac_gsurf_redclump}.
  The later evolution corresponds to the early asymptotic giant branch (AGB), followed by the thermal pulses.
  The small hook seen at $\log g \simeq 2$ corresponds to the AGB bump, which is associated to the ignition of the He-shell.
Afterwards the progressive expansion of the convective envelope leads to strong transport of AM, slowing down the core as the star goes through thermal pulses.
These pulses are seen as small loops at $\log g \simeq 0$ until the star finally leaves the AGB phase, contracts towards the white dwarf phase and enters into the cooling branch.
The track ends when the stellar luminosity drops below $\log L/L_{\odot} = -3$.
  
\begin{figure}
    \resizebox{\hsize}{!}{\includegraphics{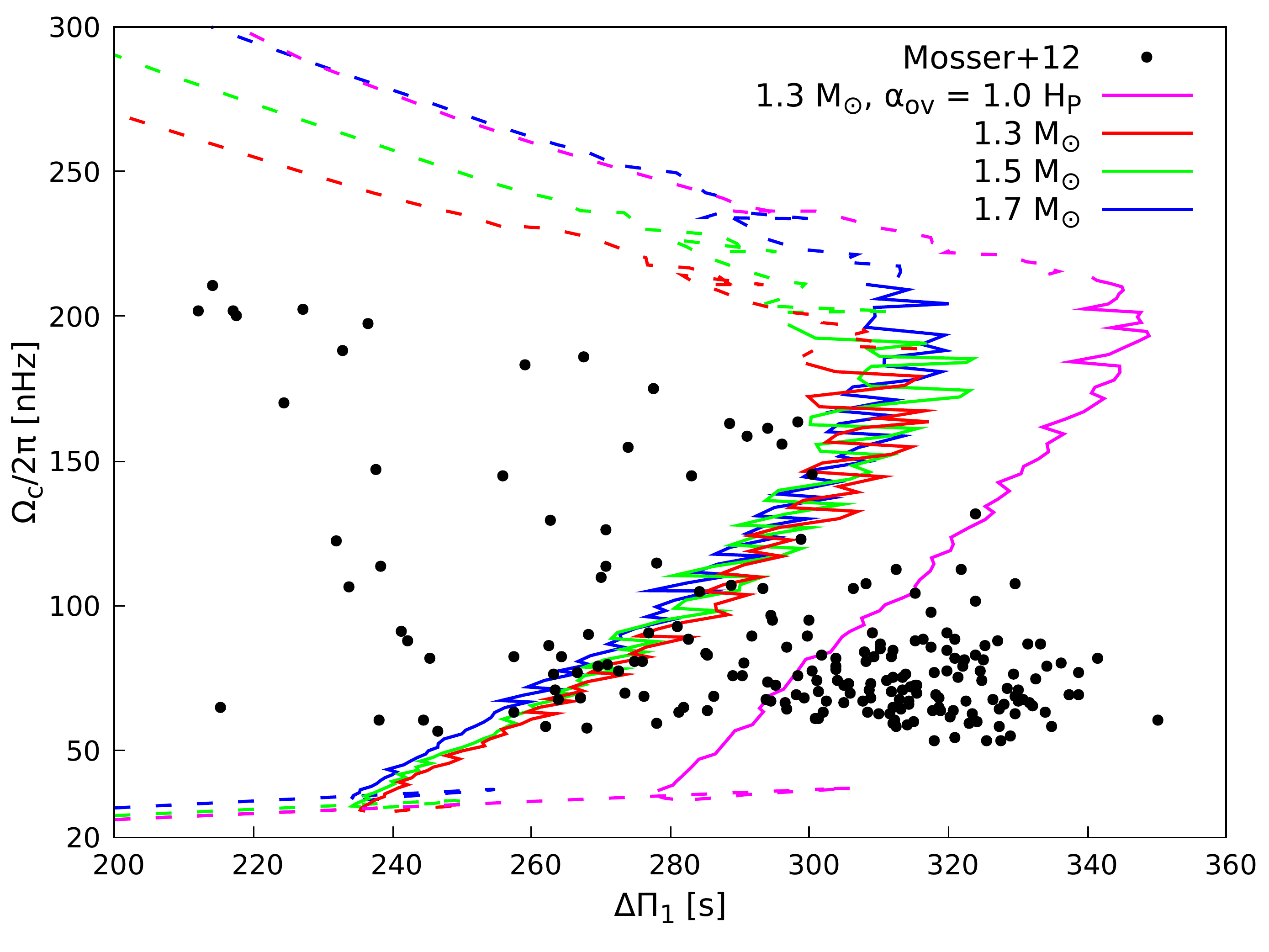}}
    \caption{Evolution of the core rotation rate during the core-helium burning phase, as a function of the asymptotic period spacing for different initial masses.
      The models correspond to those presented in Fig. \ref{omegac_gsurf_redclump}.
      The solid lines show the region where the central helium decreases from $Y=0.9$ to $Y=0.1$.
      The evolution proceeds from left to right during most of the core-helium burning phase (solid lines).
      The data points correspond to the red-clump stars presented by \citet{mosser12}.
      The magenta line shows a model where we assumed strong penetrative convection.
    }
    \label{omegac_deltapi}
  \end{figure}

 %Period spacing
 %------------------------------------------------------------------------------------------
   Another relevant point is that, in the evolutionary scenario studied here, the core of low-mass stars spins up during the core-helium burning phase with evolution.
   This is contrary to what is obtained with AM tansport by the Tayler instability \citep[e.g.][]{fuller19,eggenberger22} and offers an interesting way to distinguish between the scenario of AM transport by the parametric formulation of the AMRI presented here and AM transport by the Tayler instability.
   Indeed, a possible way to do this is to study the evolution of the core rotation rate as a function of the period spacing, since during the core-helium burning phase, the asymptotic period spacing of g-modes in stellar models increases during $\sim 80 \%$ of the core-helium burning time, because as the convective core grows the layers contributing the most to the integral move out in radius and so decreasing the value of the integral; i.e. it essentially traces the size of the convective core \citep[see Fig. 3 of][]{bossini15}.
   The asymptotic period spacing of dipole g-modes can be computed as
   \begin{equation}
     \label{eq_deltapi}
\Delta\Pi_{1} =\frac{2\pi^{2}}{\sqrt2} \left(\int_{0}^{r_{\rm g}} \rm N_{\rm BV} \frac{dr}{r}\right)^{-1}
   \end{equation}
   The evolution of the core rotation rate as a function of $\Delta\Pi_{1}$ is shown in Fig. \ref{omegac_deltapi} for models with masses $M=1.3, 1.5$ and $1.7 M_{\odot}$ during the core-helium burning phase.
   The data points correspond to red-clump stars analyzed by \citet{mosser12}.
   %--------------------------------------------------
   The noisy behaviour seen at period spacings higher than $\Delta\Pi_{1} \gtrsim 260$ s occurs because of the oscillatory nature of the mass of the convective core during this phase.
   This occurs because in our models during the core-helium burning phase the convective core grows until the radiative gradient starts developing a minimum inside and close to the edge of the convective core, which becomes lower than the adiabatic gradient and splits the convective region into two.
   Afterwards, the convective core grows until it can merge with the convective shell and the core enlarges again, repeating the cycle again until the helium is depleted \citep[see][]{salaris05}.
   With a large enough sample of stars in this phase, one could then probe the trend predicted by this scenario.
   
   We do a first comparison with the data set of \citet{mosser12}.
   Although our models can reproduce roughly the range of asymptotic spacings and core rotation rates expected, the trends in the data are not strong enough to either discard or support the scenario studied (see Fig. \ref{omegac_deltapi}).
   %NEW TEXT 4/11
   However, we do note that there is an accumulation of stars at high period spacings and low core-rotation rates, which would hint at a rather constant to decreasing core rotation rate during the core-helium burning phase.
     This is further supported by the lack of fast rotators at high period spacings, which are rather seen at low period spacings, but would probably correspond to stars that have just finished burning helium in their cores and are thus climbing the asymptotic giant branch.
   
  We also note that our standard models with an exponential overshooting extended over 0.1 $H_{\rm p}$ (red, green and blue line models in Fig. \ref{omegac_deltapi}) cannot reproduce the highest period spacings at $\Delta\Pi_{1} \gtrsim 315$ s.
     In this phase, the value of $\Delta\Pi_{1}$ is proportional to the size of the convective core \citep{montalban13}.
     Thus a possible way to reproduce the highest $\Delta\Pi_{1}$ values is to increase the degree of overshooting.
     We do this by adopting a step overshooting scheme in the convective penetration approach, that is the core is extended a distance dictated by the free parameter $\alpha_{\rm OV}$ and the region is assumed to be adiabatic (i.e. $\nabla = \nabla_{\rm AD}$).
     To reproduce the highest $\Delta\Pi_{1}$ we needed to extend the core overshooting to $\alpha_{\rm OV}= 1.0 H_{\rm P}$, which is a rather high value (for comparison on the main sequence $\alpha_{\rm OV} \simeq 0.1 - 0.2 H_{P}$, see \citet{claret16, claret18}).
     The result is shown in Fig. \ref{omegac_deltapi} in magenta lines for our $1.3 M_{\odot}$ model and is similar for different masses.
     This test shows that in our model of AM transport the discrepancy seen with the period spacing is not a product of an incorrect treatment of the overshooting during the core-helium burning phase.
     We also note that a proper treatment of convective boundary mixing to reproduce the asymptotic period spacing remains challenging \citep{bossini15}, but is out of the main scope of this work.
  Regarding the empirical data, it is still difficult to detect gravity-dominated mixed modes of stars in the upper RGB at high luminosities since the non-radial modes trapped in the core are expected to vanish due to the strong radiative damping \citep{dupret09, grosjean14} and hence the core rotation rate in this phase remains unknown.
  It is then still possible to propose a scenario where the core actually spins down during the advanced RGB since it does not pose any contradictions.

  The effect of the prescription studied in this work on the efficiency of the chemical mixing remains to be explored since in some cases, such as media with low $Pm$ and $Rm$ numbers the kinetic energy is non-negligible compared to the magnetic one \citep{gellert16}.
  We finally note that the role of chemical gradients in reducing the efficiency of AM transport by the AMRI remains to be explored.
  This is of course a key point recalling the strong inhibiting effects of chemical gradients in the case of the standard magneto-rotational instability \citep[see for instance][]{wheeler15,griffiths22} and AM transport by the Tayler instability.
  Moreover, preliminary detailed asteroseismic studies of the internal rotation of red giants suggest a possible link between the location of chemical and rotation gradients \citep{dimauro18,fellay21}.
  In this context, we found that the standard MRI does not have any impact on the core rotation rate of low-mass evolved stars, because both the thermal and chemical stratification strongly inhibits the instability, specially in regions close to the core where chemical gradients can develop (See App. \ref{smri}).
  Since it is the region from where layers tends to contract, the core then does not slow down due to this instability.

   %================================================================================  
   %================================================================================  
   \section{Conclusions}
   \label{conclusion}
   We build upon the work of \citet{spada16} by exploring the possible effects of a parametric formulation for the azimuthal magneto-rotational instability on the internal redistribution of AM for stars in different evolutionary phases.
   We demonstrate that an AM transport process whose efficiency scales with the global degree of internal shear in radiative regions can reproduce simultaneously the evolutionary trends in core rotation rate of low-mass stars in the H-shell burning phase and in the core-helium burning phase as given by asteroseismic constraints \citep[e.g.][]{mosser12,gehan18}.
   Remarkably, in our models the core spins up during the core-helium burning phase, which is particular of this kind of parametrisation, where the cores spin down considerably in the upper part of the RGB; still it remains a possible scenario which remains to be confirmed.
   
   Using the same parametrisation for the AM transport in intermediate-mass stars, the efficiency with respect to low-mass stars needs to be enhanced in order to correctly reproduce the rotational constraints of red giants in the H-shell burning phase.
   In this direction, we explored a prescription that takes into account the enhancement of the AM transport by the molecular viscosity, considering its maximum value in the radiative regions and find that in our intermediate-mass models the core rotation rate of both H-shell and core helium burning stars can be better reproduced.

  Since possibly a process that acts more strongly as the core decouples from the envelope can reproduce the core rotation rate of red giants in the hydrogen-shell and core-helium burning phases simultaneously, we suggest this process should be further explored with direct numerical simulations under different rotational configurations.
  In particular, the critical role of chemical gradients on the inhibition of the AMRI should be studied.
  If this instability is not strongly inhibited by chemical gradients, it could constitute a missing piece of the AM transport puzzle in both low- and intermediate-mass stars.
   
  \begin{acknowledgements}
    FDM and PE have received funding from the European Research Council (ERC) under the European Union’s Horizon 2020 research and innovation programme (grant agreement No. 833925, project STAREX).
    FS is supported by the German space agency (Deutsches Zentrum f\"ur Luft- und Raumfahrt) under PLATO data grant 50OO1501.
    FDM thanks Georges Meynet, Sébastien Salmon, Gaël Buldgen, Thibaut Dumont and Adolfo Simaz-Bunzel for useful discussions at different stages of this work.
   We thank the anonymous referee for her/his constructive input, which improved the outline of this article.
  \end{acknowledgements}

%==========================================================================
\bibliographystyle{aa}
\bibliography{visc_amri}
%==========================================================================
%===========APENDICES===========
%==========================================================================
  \begin{appendix}
  %=============================================================================================
  \section{Standard Magneto-rotational instability}
  \label{smri}
\begin{figure}
    \resizebox{\hsize}{!}{\includegraphics{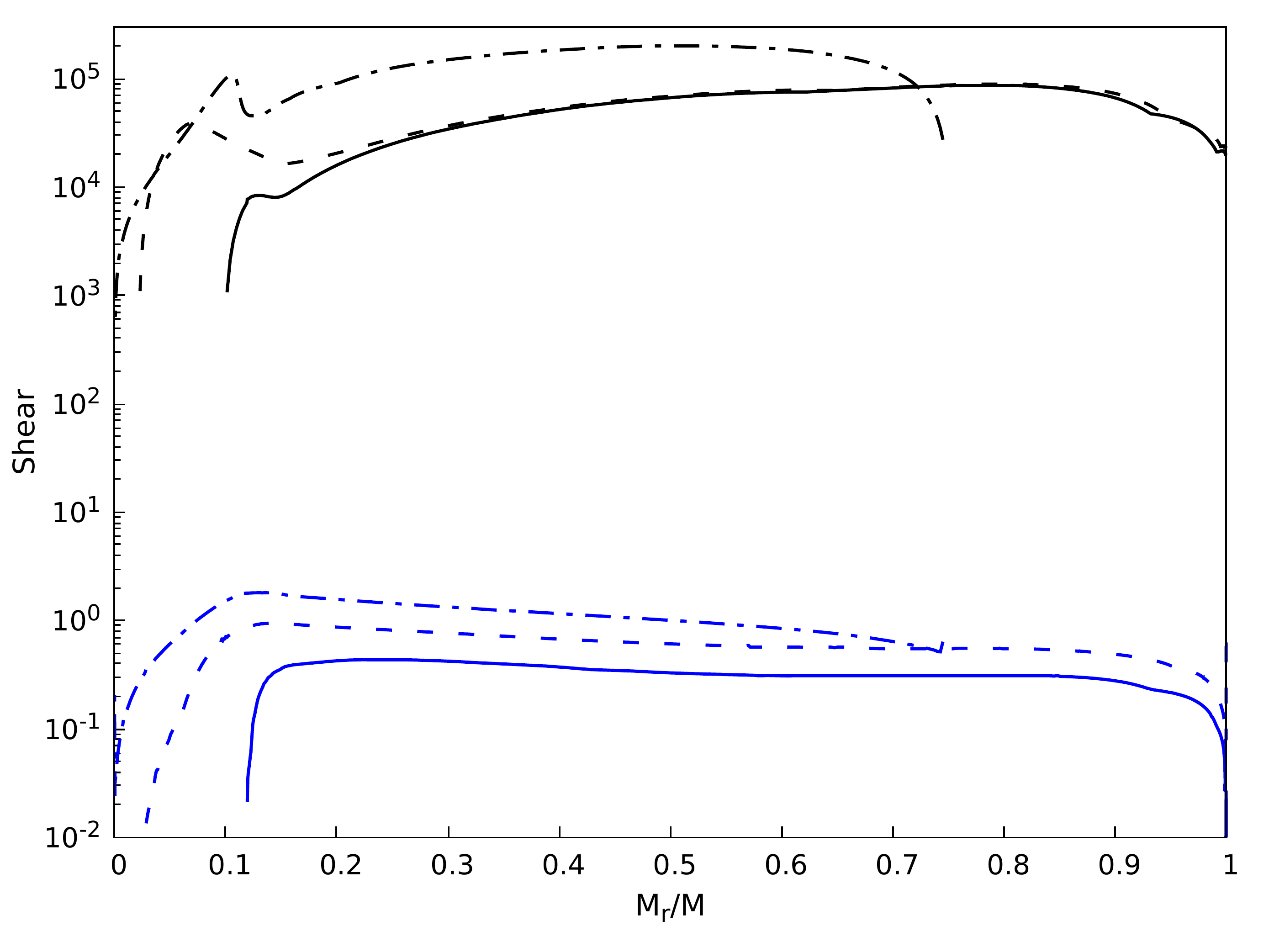}}
    \caption{Local shear as a function of the mass coordinate for a $1.7 M_{\odot}$ during the main sequence when the central hydrogen mass fraction is $X=0.35$ (solid lines), at the end of the main sequence (dashed lines), and at the RGB base (dot-dashed lines).
      The black lines correspond to the minimum shear required to trigger the standard MRI, while the blue line is the actual shear of the model.
      The regions where the line is missing correspond to either the convective core or the convective envelope.
    }
    \label{qmin_mri}
    \end{figure}

  In our models the standard magneto-rotational instability (MRI) does not transport angular momentum because both thermal and chemical stratification stop the instability from triggering.
    The criterion to trigger the instability is set by the minimum local shear needed, defined as $q_{\rm min}= - 2 (N_{\rm eff}/\Omega)^{2}$ \citep[e.g.][]{wheeler15} where $q \equiv \partial\ln\Omega / \partial\ln r$ is the shear, $N_{\rm eff}$ is the effective Brunt-V\"ais\"al\"a frequency and $\Omega$ the angular velocity.
    As long as $q < q_{\rm min}$ the MRI does not trigger, which is the case in our low-mass models.
    This is illustrated in Fig. \ref{qmin_mri} for a model computed only with hydrodynamical processes.
    The minimum shear $q_{\rm min}$ remains orders of magnitude higher than the actual shear $q$, showing so that the standard MRI is not able to transport angular momentum through evolution in low-mass stars.

\end{appendix}
%==========================================================================
\end{document}